\documentclass[useAMS, usegraphicx]{mn2e}
\title[Radio Monitoring of 3C~84 in the Period of 2009-2011]{VLBI and Single Dish Monitoring of 3C~84 in the Period of 2009-2011}
\author[H. Nagai et al.]{H. Nagai, $^{1}$\thanks{E-mail:hiroshi.nagai@nao.ac.jp} M. Orienti,$^{2, 3}$ M. Kino,$^{1}$ K. Suzuki,$^{4}$ G. Giovannini,$^{2, 3}$ A. Doi,$^{5}$ \newauthor K. Asada,$^{6}$ M. Giroletti,$^{2}$ J. Kataoka,$^{7}$ F. D'Ammando,$^{2}$ M. Inoue,$^{6}$ A. L\"ahteenm\"aki,$^{9}$ \newauthor M. Tornikoski,$^{9}$ J. Le\'on-Tavares,$^{9}$ S. Kameno,$^{8}$ U. Bach$^{10}$
\\
$^{1}$ National Astronomical Observatory of Japan, Osawa 2-21-1, Mitaka, Tokyo 181-8588, Japan\\
$^{2}$ INAF Istituto di Radioastronomia, via Gobetti 101, 40129, Bologna, Italy\\
$^{3}$ Dipartimento di Astronomia, Universita' di Bologna, via Ranzani 1, I-40127, Bologna, Italy \\
$^{4}$ Department of Astronomy, Graduate School of Science, The University of Tokyo, 7-3-1 Hongo, Bunkyo-ku, Tokyo 113-0033, Japan\\
$^{5}$ Institute of Space and Astronautical Science, Japan Aerospace Exploration Agency, 3-1-1 Yoshinodai, Chuo-ku, Sagamihara,\\ Kanagawa 229-8510, Japan\\
$^{6}$ Institute of Astronomy and Astrophysics, Academia Sinica. P.O. Box 23-141, Taipei 10617, Taiwan, R.O.C.\\
$^{7}$ Research Institute for Science and Engineering, Waseda University, 3-4-1, Okubo, Shinjuku, Tokyo, 169-8555, Japan\\
$^{8}$ Faculty of Science, Kagoshima University, 1-21-35 Korimoto, Kagoshima 890-0065, Japan\\
$^{9}$ Aalto University  Mets\"{a}hovi Radio Observatory,  Mets\"{a}hovintie 114, FIN-02540 Kylm\"al\"a, Finland \\
$^{10}$ Max-Planck-Institut f\"{u}r Radioastronomie, Auf dem H\"{u}gel 69, 53121 Bonn, Germany }

\begin{document}
\date{ }

\pagerange{\pageref{firstpage}--\pageref{lastpage}} \pubyear{2002}

\maketitle

\label{firstpage}

\begin{abstract}
The radio galaxy 3C~84 is a representative of $\gamma$-ray-bright misaligned active galactic nuclei (AGNs) and one of the best laboratories to study the radio properties of the sub-pc jet in connection with the $\gamma$-ray emission.  In order to identify possible radio counterparts of the $\gamma$-ray emissions in 3C~84, we study the change in structure within the central 1~pc and the light curve of sub-pc-size components C1, C2, and C3.  We search for any correlation between changes in the radio components and the $\gamma$-ray flares by making use of VLBI and single dish data.  Throughout the radio monitoring spanning over two GeV $\gamma$-ray flares detected by the {\it Fermi}-LAT and the MAGIC Cherenkov Telescope in the periods of 2009 April to May and 2010 June to August, total flux density in radio band increases on average.  This flux increase mostly originates in C3.  Although the $\gamma$-ray flares span on the timescale of days to weeks, no clear correlation with the radio light curve on this timescale is found.  Any new prominent components and change in morphology associated with the $\gamma$-ray flares are not found on the VLBI images.
\end{abstract}


\section{Introduction}
With the Large Area Telescope (LAT) on board {\it Fermi Gamma-ray Space Telescope}, GeV $\gamma$-ray emission from 3C~84 (alias NGC~1275, z=0.0176) was detected for the first time (Abdo et al. 2009).  It was one of the two misaligned AGNs detected by {\it Fermi} after four-months operations.  More recently, $\gamma$-ray emission is detected in 12 misaligned AGNs (Abdo et al. 2010a, Brown \& Adams 2012).  This gives new challenges to the AGN unification scenario.  Among these misaligned AGNs, 3C~84 exhibited strong variability in its $\gamma$-ray emission.  An averaged $\gamma$-ray flux during the first four months is $(2.10\pm0.23)\times10^{-7}$~ph cm$^{-2}$ s$^{-1}$ above 100~MeV.  This $\gamma$-ray flux is seven times brighter than the upper limit by {\it EGRET}.  It was claimed that the innermost jet of 3C~84 is the most likely source of $\gamma$-ray emission because of the time variation on the timescales shorter than years to decades (Abdo et al. 2009).  During the first 2 years, {\it Fermi}-LAT detected $\gamma$-ray emission from 3C~84, two episodes of increased $\gamma$-ray activity were observed: one occurred in 2009 Apr-May (Kataoka et al. 2010), and the other one occurred in 2010 June-August (Brown \& Adams 2011).  In particular, the second flare was detected up to 102.5~GeV and very-high-energy $\gamma$-ray emission with a spectral cut-off around 500~GeV was found successively by MAGIC (Aleksi{\'c} et al. 2011).
	
The radio source 3C~84 is known to be a bright radio source, and has been studied extensively since the early days of radio astronomy.  The radio flux has been monitored since 1960, and episodes of violent flux increase were reported (Kellermann \& Pauliny-Toth. 1968; O'Dea et al. 1984).  In mid-1980s, the radio flux became exceptionally bright, more than 60~Jy at centimeter wavelengths, and then subsequently decreased such that, by the early 2000s, the radio flux decreased to $\sim10$~Jy (Nagai et al. 2010, hereafter Paper I).  This flux decrease can be ascribed to the adiabatic expansion of radio lobe in the central $\sim5$~pc (Asada et al. 2006).  The viewing angle of the jet is estimated to be 11-55$^{\circ}$ to the line of sight (e.g., Asada et al. 2006; Lister et al. 2009).   Around 2005, it was reported that the radio flux started to increase again (Abdo et al. 2009).  The VLBI observations revealed that this flare originated from within the central pc-scale core, accompanying the ejection of a new jet component (Paper I).  This new component appeared from the south of the core around 2003 (Suzuki et al. 2012, hereafter Paper II), and was moving to the position angle $\sim160^{\circ}$ steadily with slightly changing speed in both parallel and perpendicular directions.  

In Paper I, we revealed that the central 1-pc structure mainly consists of three bright components C1, C2, and C3.  In Paper II, we argued that it is difficult to reconcile the observed GeV $\gamma$-ray emission with a one-zoned synchrotron-self Compton model in C3 because the observed spectral index of C3 between 22 and 43~GHz disagrees with that derived from the SED modeling.  In order to obtain an additional validation to this argument, this paper focuses on the comparison between radio light curve and GeV $\gamma$-ray flares which occurred in the period of 2009 April-May and 2010 June-August.  We argue which component is the most likely site of the observed $\gamma$-ray emission by comparing radio flux variability with the $\gamma$-ray flaring events.  We are also interested in the possible emergence of a new radio component associated with the $\gamma$-ray flares.  

In this paper, we present the light curve of each component obtained by the Monitoring Of Jets in Active galactic nuclei with VLBA Experiments\footnote{The MOJAVE data archive is maintained at http:\slash\slash www.physics.purdue.edu\slash MOJAVE.} (MOJAVE: Lister et al. 2009) and detailed structural change obtained by the VLBI Exploration Radio Astrometry (VERA).  Moreover, the 37-GHz data from the Mets\"{a}hovi have been used to complement the MOJAVE light curve.  Throughout this paper, we adopt $H_{0}=70.2 $~km sec$^{-1}$ Mpc$^{-1}$, $\Omega_{\mathrm{M}}=0.27$, and $\Omega_{\mathrm{\Lambda}}=0.73$ (Komatsu et al. 2011; 1~mas=0.35~pc, 0.1~mas~yr$^{-1}=0.11c$). 

\section{Observation and Data Reduction}
\subsection{The VERA data}\label{sect:[VERA-DATA]}
Observations were carried out between 2009 April and 2011 March for 8 epochs with four VERA stations at 43~GHz.  Each observation lasted 8 hours, consisting of a number of scans in which the length of each scan is 570 seconds, with a 30-seconds gap inserted between each scan.  Most of the scans were assigned for 3C~84 and a few scans were assigned for calibration.  The total on-source time for 3C~84 is about 7 hours for each observation.  The data were recorded at a rate of 128 Mbps for the first 6 epochs and 1024~Mbps for the remaining 2 epochs, with 16-MHz and 128-MHz bandwidth, respectively.  

Data reduction was performed using the NRAO Astronomical Imaging Processing System ({\sc AIPS}).  A standard $a$ $priori$ amplitude calibration was performed using the {\sc AIPS} task {\sc APCAL} based on the measurements of the system temperature during the observations and the aperture efficiency provided by each station for VERA data at 43~GHz.  Fringe fitting was done using the {\sc AIPS} task {\sc FRING}.  For the deconvolution of synthesized image, we used CLEAN and self-calibration technique.  Final images were obtained after a number of iterations with CLEAN and both phase and amplitude self-calibration using the {\sc DIFMAP} software package (Shepherd 1997). 
 
After the process described above, we found that the VERA 43-GHz was resolved and some portion of flux density within the central $\sim1$~pc region was missing.  Because of this missing flux, it was difficult to argue the flux change using VERA. 
We do not use the VERA data for the argument of light curve, but they are used to study the structural change because the central region is less opaque at 43~GHz than at 15~GHz, i.e., the frequency of the MOJAVE data. 

\subsection{The MOJAVE data}
We imported the calibrated {\it uv}-data of the MOJAVE programme (15~GHz) into the NRAO {\sc AIPS} package and we performed a few phase-only self-calibration iterations and flagging some data when the amplitudes were too high or too low with respect to the other visibilities. Finally we produced high- and low-resolution images in order to study either the core structure or the extended emission. The {\it high-resolution} 
image (FWHM=0.55 mas) was produced considering only the baselines longer than 50 M$\lambda$ and using pure uniform weight, whereas the {\it low-resolution} image was obtained using the baseline shorter than 250 M$\lambda$ and natural weight (FWHM=1~mas). 

\subsection{The Mets\"{a}hovi data}
The single dish monitoring data at 37 GHz is adopted from the Mets\"{a}hovi quasar monitoring program. The observations were carried out with the Mets\"{a}hovi 14-m radio telescope.  A detailed description of the data reduction and analysis is given in Ter\"{a}sranta et al. (1998). In order to compare the VLBI and single dish light curves, we selected the data from 2009 January 1 to 2011 November. 

\section{Model Fits}\label{sect[modelfit]}
\begin{figure*}
\begin{center}
\includegraphics[width=16.5cm]{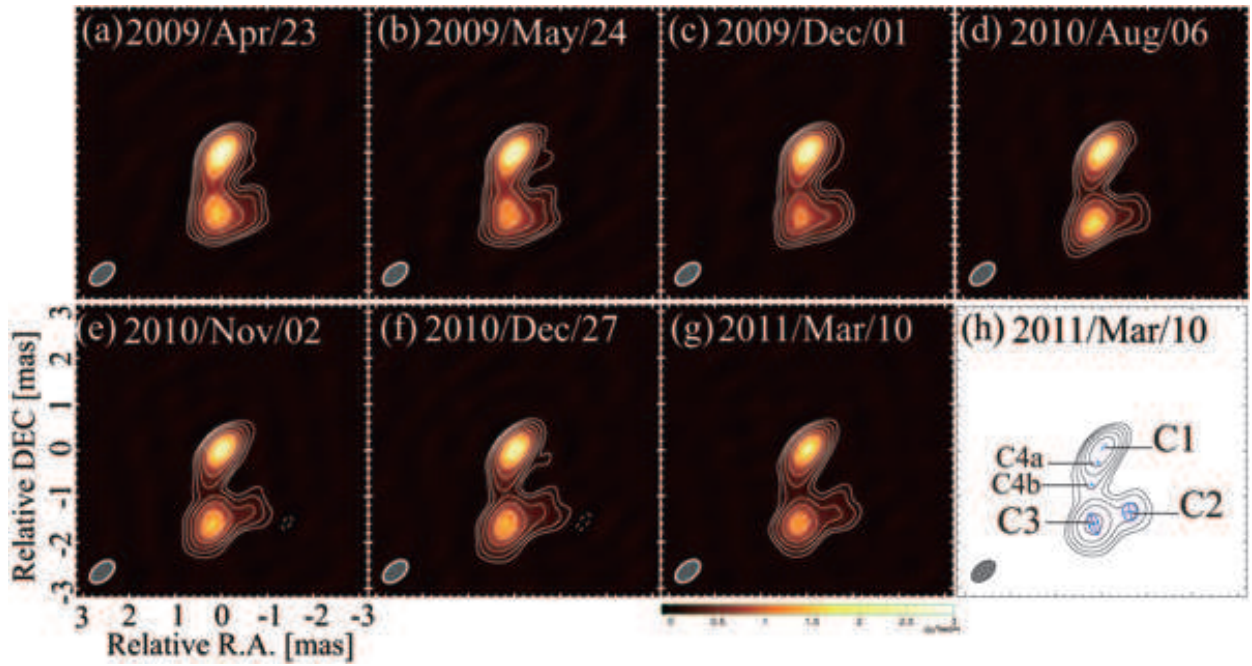}
\end{center}
\caption{(a)-((g) VERA 43-GHz images and model fit image.  The ellipse indicated at the bottom left corner in each image is the restoring beam.  The size of the restoring beam is ($0.63\times0.37$)~mas at the position angle of $-50^{\circ}$, which is a typical beam size of all observing epochs.  The contours are plotted at the level of (-1, 1, 2, 4, 8, 16, 32) $\times$ 62.6~mJy/beam, which is the 3-times image noise rms on 2010 December 27 (f).  (h) Model-fit image on 2011 March 10.  Symbols indicated by blue lines are the model components obtained by the Gaussian model fit.  The plus signs are point sources (C4a and C4b) which are placed to model the emission between C1 and C3 (see text in \S\ref{sect[modelfit]}).}
\label{fig:[IMAP-MODEL]}
\end{figure*}

In order to investigate if there is any structural change, i.e., new component emergence, we deconvolved the VERA 43-GHz images by using the several components whose brightness distribution is elliptical Gaussian. 
Throughout all epochs for VERA data at 43~GHz, overall structures were mostly well-represented by three major components C1, C2, and C3, which were the same components identified in Paper I and Paper II.  In addition to these three components, there was an additional emission bridging between C1 and C3.  We had a difficulty to model this emission: relative flux densities and positions of the Gaussian components in this region can be correlated with those of C1 and C3, and the position and size of C1 and C3 were often skewed by these components.  We operationally fitted this bridging emission by using one or two point sources (unresolved components) instead of Gaussians (see C4a and C4b in Figure \ref{fig:[IMAP-MODEL]}), in order to avoid them from such a skewing.  The choice of one or two point sources was somewhat arbitrary, and it was difficult to keep consistency across the epochs.  We always tried to maintain consistency of the three main components (C1, C2, and C3) even if the position and flux of the point sources C4a and C4b did not show smooth change across the epochs.  
We do not use the result from this model fit for the evaluation of flux variability because of significant missing flux which is mentioned in section \ref{sect:[VERA-DATA]}, but we use the VERA data for the discussion of structural change.        
Figure \ref{fig:[IMAP-MODEL]} shows the CLEAN images from the VERA and one example of model-fit image.  Note that the model-fit image and corresponding CLEAN image are essentially similar, demonstrating the validity of modeling procedure. 

The analysis of the 15-GHz MOJAVE data was performed on the image domain instead of on the visibility domain.  The flux density, size and position of each component was derived by means of the AIPS task JMFIT, which performs a Gaussian fit to the components on the image plane.  This approach has been found preferable rather than the visibility-based approach due to the presence of extended structures. 
Indeed, the minimum baseline of the VLBA at 15 GHz (6.5 M$\lambda$) is sensitive to the extended flux density arising from both the northern and southern lobes located about 15~mas and 10~mas from the core, respectively (e.g., Asada et al. 2006).  Both components had complex sub-structures that make the analysis of the visibility data difficult to carry out.  On the other hand, the minimum baseline of VERA was too long to detect such extended structures, making the model fitting of the core components on the visibility data more reliable.  We use the result from MOJAVE data for the discussion of the light curve.

\section{Structure and Light curve}\label{sect[Results]}
\begin{figure}
\begin{center}
\includegraphics[width=8cm]{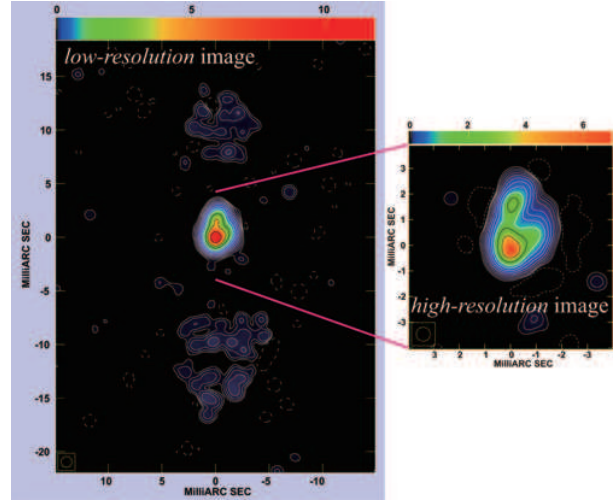}
\end{center}
\caption{Left:{\it Low-resolution} image of MOJAVE 15~GHz on 2011 December 12.  The size of restoring beam is ($1\times1$)~mas.  The contours are plotted at the level of (-1, 1, 2, 4, 8, 16, 32, 64, 128, 256, 512, 1024)$\times$3.8~mJy/beam, which is the 3-times image noise rms.  Right:{\it High-resolution} image of the central ($8\times8$)~mas region on the same date.  The contour levels are same with those in {\it low-resolution} image.}
\label{fig:[MOJAVE]}
\end{figure}

Figures \ref{fig:[IMAP-MODEL]} and \ref{fig:[MOJAVE]} show the total intensity images of MOJAVE at 15~GHz and VERA at 43~GHz, respectively.  Two bright components C1 and C3 are separated by about 2 milli-arcsec in the north-south direction, while the fainter component C2 is located on the west of C3 (see Figure \ref{fig:[IMAP-MODEL]}).  The overall structure is very similar to that presented by Paper I and Paper II.  From the analysis of the data presented in this paper, no significant changes in the central structure of 3C~84 has been detected.  C3 is continuously moving to the south with a position angle $\sim170^{\circ}$ with respect to C1, as it has already been found in previous papers.  The positional change of C2 is somewhat random rather than systematic, but it seems that C2 is also moving to the south slightly on average.  A detailed analysis of the kinematics of the central components of 3C~84 will be discussed in a forthcoming paper. 

On average the total flux density increases with time, but some rise and fall in flux density on the timescales of a month is also seen in the Mets\"{a}hovi data (Figure \ref{fig:lightcurve}).  Total flux increase shows similar trend with the sum of flux of C1, C2, and C3.  The most of flux increase originates in C3.  The increasing level is more than a factor of two throughout the monitoring (see Table \ref{tab:MOJAVE15}).  On the other hand, the flux variation of C2 and C1 is only about 10\%.  In Paper II it was found that C3 showed moderate flux increase until 2008 and then both C1 and C3 showed a rapid flux increase after 2008.  Such a rapid flux increase of C1 is not seen anymore.  Surprisingly, any new component ejection accompanied by this flux increase is not detected with the exception of minor components C4a and C4b.  This is discussed in the section \ref{sect[gamma-radio]}. 

\section{Comparison with $\gamma$-ray Events and radio light curve}\label{sect[gamma-radio]}
\begin{figure*}
  \begin{center}
    \includegraphics[width=17cm]{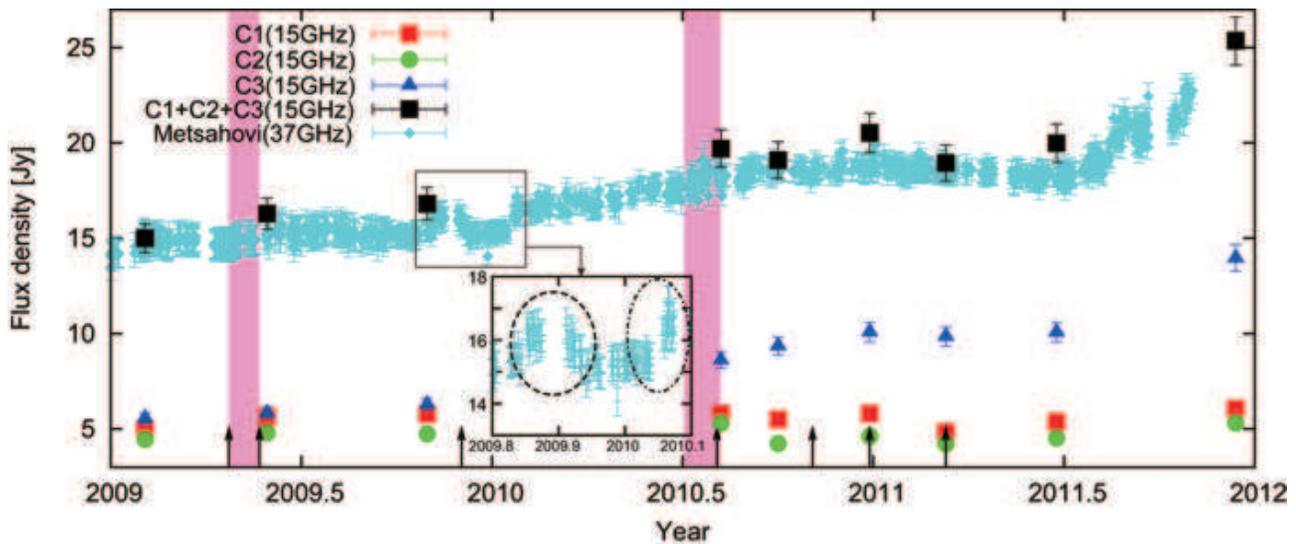}
  \end{center}
  \caption{Single dish and VLBI light curves from Metsahovi at 37~GHz and MOJAVE at 15~GHz, respecctively.  Red squares, green circles, and blue triangles represent the MOJAVE flux densities of C1, C2, and C3, respectively.  The sum of flux densities of C1, C2, and C3 is indicated by black squares.  Single dish light curve by Mets\"{a}hovi is indicated by cyan diamonds.  Two $\gamma$-ray flaring events are indicated by pink vertical lines.  Vertical arrows represent the date of VERA observations.  Small graph inserted in overall plot is the Mets\"{a}hovi light curve from 2009.8 to 2010 where the range indicated by the square of thin solid line (see text in \S\ref{sect[gamma-radio]}).  Note that Mets\"{a}hovi flux density contains the emission from the entire 3C~84 source.}\label{fig:lightcurve}
\end{figure*}

In Figure \ref{fig:lightcurve}, we present the radio data collected by Mets\"{a}hovi and MOJAVE.  The pink vertical lines indicate the $\gamma$-ray flaring periods reported by Kataoka et al. (2010) and Brown \& Adams (2011).  The first one occurred in the period of 2009 April-May.  The flaring timescale is about 1 month.  The MOJAVE light curve shows that the flux density of C1 obtained in 15 days after the peak of this flare ($4.98\pm0.25$~Jy on 2009 May 28) increases slightly as compared to that obtained in $\sim4$ month before ($5.65\pm0.28$~Jy on 2009 January 30).  The flux density of C3 could be also increasing between 2009 January 30 ($5.56\pm0.28$~Jy) and 2009 May 28 ($5.84\pm0.29$~Jy), but the increasing level is not significant as compared to the flux calibration accuracy of VLBA ($\sim5$\%).   
On the other hand, no obvious change in radio flux density is found from the Mets\"{a}hovi 37-GHz data on the timescale of the $\gamma$-ray flaring event.  No structural change on the VLBI image is also seen before and after the flare despite our VERA observations pin down right beginning and end of the flare (2009 April 23 and 2009 May 24).  

The second $\gamma$-ray flare occurred in the period of 2010 June-August.  Its e-folding rise time is about 1.5 days and a subsequent e-folding decay time is about 2.5 days.  This flare occurred during the increasing activity of total flux density at 37~GHz spanning from 2010 to 2011.  From Figure \ref{fig:lightcurve} it is found that this radio flux increase mostly originates in C3.  However, by focusing on the timescale of days, the light curve of total flux does not show any signature of flare.  No obvious change in VLBI-scale structure before and after the flare is seen.    

In the period of both $\gamma$-ray flares, no obvious change in radio flux density is detected by Mets\"{a}hovi on the timescale of the $\gamma$-ray flaring events (days to weeks).  If the variability in the radio band had been of the same magnitude as in the $\gamma$-rays, where the flux doubled or tripled its value, the single dish observations would have detected such variation.  Thus, we may conclude that the magnitude of radio variability is smaller than that in $\gamma$-ray band or the size of the radio counterpart is so small that the radio variability of the flaring component is mitigated by the constant activity of the other components.  However, we must note that Mets\"{a}hovi 37-GHz light curve shows a small rise and subsequent decay in the period from 2009 October to 2009 December (indicated by the oval of broken line in the small graph in Figure 3).  The timescale of this change is about one month.  Right after this rise and decay, a small increasing activity which occurs on the timescale of one week is detected again (indicated by the oval of broken dotted line in the small graph in Figure 3).  Either of these changes in the 37-GHz light curve is possibly related to the $\gamma$-ray flare with some time delay, as proposed in Le\'{o}n-Tavares et al. (2011).  More recently, an increasing of radio activity is observed from around 2011 June.  This probably originates in C3 because the flux density of C3 increases between 2011 June 24 and 2011 December 12 with the significance of 4.5$\sigma$ (Table \ref{tab:MOJAVE15}).  This increasing activity seems to be difficult to relate to the $\gamma$-ray flare because the time difference between the two events is very large ($\sim1$ year).  Further multifrequency analysis to solve this puzzle will be presented in a forthcoming paper. 

Where is the radio counterpart of the $\gamma$-ray flares?  One possibility is that the $\gamma$-ray flaring region is embedded in the self-absorbed core, i.e., C1 or further upstream of jet.  This seems quite natural because the $\gamma$-ray time variation is very fast, indicating a small emission region size.  In particular, the resultant emission region size for the case of second flare is estimated to be $\sim10^{-3}$~pc if we assume a Doppler factor ($\delta$) of 2 (Brown \& Adams 2010).  The size of $\sim10^{-3}$~pc is smaller than the beam size of VLBA at 15~GHz by two orders of magnitude.  Therefore, it is possible that time variation of radio counterpart is mitigated by the contamination from other radio emitting region.  However, a new jet component associated with the $\gamma$-ray flare is likely to be ejected, and such a component may be visible on the image some time after the $\gamma$-ray flare as a result of propagation down to optically thin region.  If the jet components are ejected from the core accompanied by the first and the second $\gamma$-ray flares with $\delta=2$, they should be visible on the image between C1 and C3 during our monitoring period.  As we mentioned in section \ref{sect[modelfit]}, there is a smoothly distributed radio emission between C1 and C3.  This emission can be modeled with two point sources.  They might be jet components accompanied by the $\gamma$-ray flares.  

It is worth pointing out for comparison that not all the $\gamma$-ray flare detected so far in blazars have a clear radio counterpart, as in the case of the GeV-TeV $\gamma$-ray flare in 3C~279 (Abdo et al. 2010b).  Another intriguing case in PKS~1510-089 is that some $\gamma$-ray flares seem to be related to changes in radio band (e.g., Marscher et al. 2010; Orienti et al. 2011) while others show no relation (e.g., D'Ammando et al. 2009).  Therefore, the lack of significant changes in radio band for 3C~84 after the detection of high $\gamma$-ray activity leaves the debate on the region responsible for the high-energy emission and its location still open.  
	
\begin{table}
  \caption{Flux densities of the central components from MOJAVE 15-GHz data}\label{tab:MOJAVE15}
  \begin{center}
    \begin{tabular}{cccc} \hline\hline
Epoch & C1 [Jy] & C2 [Jy] & C3 [Jy] \\ \hline
2009/Jan/30　& 4.99$\pm$0.25 & 4.46$\pm$0.22 & 5.56$\pm$0.28 　\\
2009/May/28　& 5.65$\pm$0.28 & 4.80$\pm$0.24 & 5.84$\pm$0.29 　\\
2009/Oct/27　& 5.78$\pm$0.29 & 4.74$\pm$0.24 & 6.30$\pm$0.31 　\\
2010/Aug/6　& 5.80$\pm$0.29 & 5.30$\pm$0.26 & 8.61$\pm$0.43 　\\
2010/Sep/29　& 5.51$\pm$0.28 & 4.24$\pm$0.21 & 9.35$\pm$0.47 　\\
2010/Nov/20　& 5.81$\pm$0.29 & 4.63$\pm$0.23 & 10.09$\pm$0.50 　\\
2011/Feb/11　& 4.84$\pm$0.24 & 4.22$\pm$0.21 & 9.85$\pm$0.49 　\\ \hline
\end{tabular}
  \end{center}
 \end{table}

\section*{Acknowledgments}
We thank Anthony Brown, the referee, for careful reading and helpful comments.  Part of this work was done with the contribution of the Italian Ministry of Foreign Affairs and University and Research for the collaboration project between Italy and Japan.  The VERA is operated by the National Astronomical Observatory of Japan.  This research has made use of data from the MOJAVE database that is maintained by the MOJAVE team (Lister et al. 2009).  The VLBA is operated by the US National Radio Astronomy Observatory (NRAO), a facility of the National Science Foundation operated under cooperative agreement by Associated Universities, Inc.  This work has made use of observations obtained with the 14 m Mets\"{a}hovi Radio Observatory, a separate research institute of the Aalto University School of Electrical Engineering. The Mets\"{a}hovi team acknowledges the support from the Academy of Finland to our observing projects (numbers 212656, 210338, 121148, and others).

\newcommand{\aap}{A\&A}
\newcommand{\apj}{ApJ}
\newcommand{\apjl}{ApJL}
\newcommand{\apjs }{ApJS}
\newcommand{\aj}{AJ}
\newcommand{\pasj}{PASJ}
\newcommand{\mnras}{MNRAS}
\newcommand{\aaps}{A\&AS}

\end{document}